\documentclass[aps,prd,preprint,preprintnumbers]{revtex4}

\usepackage{amsmath}
\usepackage{graphicx}

\begin{document}

\preprint{MPP-2009-3}

\title{Unparticle physics and neutrino phenomenology}

\author{J. Barranco}
\email{jbarranc@aei.mpg.de}

\affiliation{Max-Planck-Institut f\"ur Gravitationsphysik 
(Albert-Einstein-Institut), Am M\"uhlenberg 1, D-14476 Golm, Germany }

\author{A. Bola\~nos}
\email{azucena@fis.cinvestav.mx}

\author{O. G. Miranda}
\email{Omar.Miranda@fis.cinvestav.mx}

\affiliation{Departamento de F\'{\i}sica, Centro de Investigaci{\'o}n y de
  Estudios Avanzados del IPN, Apartado Postal 14-740 07000
  M\'exico, D F, M\'exico}

\author{C. A. Moura}
\email{moura@na.infn.it}

\affiliation{INFN Sezione di Napoli, Complesso Universitario Monte S. Angelo,
  Via Cintia, I-80126 Napoli, Italy}

\altaffiliation[Also at ]{Instituto de F\'{\i}sica Gleb Wataghin
  - UNICAMP, 13083-970 Campinas, SP Brazil}

\author{T. I. Rashba}
\email{timur@mppmu.mpg.de}

\affiliation{Max-Planck-Institut f\"ur
Physik (Werner-Heisenberg-Institut), F\"ohringer Ring 6, 80805
M\"unchen, Germany}

\affiliation{Pushkov Institute of Terrestrial Magnetism,
  Ionosphere and Radio Wave Propagation of the Russian Academy of Sciences,
  142190, Troitsk, Moscow region, Russia}

\altaffiliation[Present address:  ]{Max-Planck-Institute for Solar System
Research, Katlenburg-Lindau, 37191, Germany}


\begin{abstract}  
  We have constrained unparticle interactions with neutrinos and
  electrons using available data on neutrino-electron elastic
  scattering and the four LEP experiments data on mono photon
  production. 
  We have found that, for neutrino-electron elastic scattering, the
  MUNU experiment gives better constraints than previous reported
  limits in the region $d>1.5$.  The results are compared with the
  current astrophysical limits, pointing out the cases where these
  limits may or may not apply.
  We also discuss the sensitivity of future experiments to unparticle physics. 
  In particular, we show that the
  measurement of coherent reactor neutrino scattering off nuclei could
  provide a good sensitivity to the couplings of unparticle
  interaction with neutrinos and quarks. We also discuss the case 
  of future neutrino-electron experiments as well as the International Linear 
  Collider. 
\end{abstract}


\maketitle

\section{Introduction}

Motivated by the idea that a scale invariant sector could exist above
TeV energies and could be probed at present or future colliders, it
has been proposed~\cite{Georgi:2007ek,Georgi:2007si} a scenario where
it is possible to calculate the interaction of such a sector with the
Standard Model (SM) sector in the low energy limit. In this case, with
the help of effective field theory, in particular with Banks-Zaks
fields~\cite{Banks:1982gt}, it is possible to obtain quantitative
results. In this limit, the scale invariant sector with scale
dimension $d$ looks like a nonintegral number $d$ of invisible
particles, named unparticles~\cite{Georgi:2007ek}.

From the phenomenological point of view, it is interesting that the
low-energy processes involving unparticles can have a particular
energy spectrum, that is not predicted by other types of new physics.
There is a rich phenomenology that can be extracted from the
unparticle idea and currently there are several constraints on the
relevant parameters of unparticle physics using a wide variety of
processes: collider phenomenology, flavour physics, top quark physics,
Higgs physics, supersymmetry, dark matter, etc. (for a recent review
see, e.g.,~\cite{Cheung:2008xu} and also, for more recent works, 
Ref.~\cite{recent}).

On the other hand, measurements of neutrino elastic scattering off
leptons and quarks are becoming more and more precise and provide a
sensitive tool to probe Neutrino Nonstandard Interactions (NSI)
and various kinds of new physics beyond the SM.  For
example, new limits on the nonstandard neutrino-electron
couplings~\cite{Barranco:2005ps,Barranco:2007ej} and on the neutrino
charge radius~\cite{Barranco:2007ea} from all neutrino-electron
scattering experiments have been recently derived.
As for nuclei the sensitivity of future low energy coherent
neutrino-nucleus scattering experiments to NSI neutrino-quark
interactions has also been studied in
detail~\cite{Barranco:2005yy,Barranco:2007tz,Barranco:2008tk}.

Neutrino data can offer the possibility of studying unparticle
phenomenology in two ways: first by effects of virtual unparticles
exchanged between fermionic currents, second by the direct production
of unparticles.  The neutrino-electron and neutrino-nuclei scattering
are examples where unparticle effects of the first type are
measurable, while single-photon production ($e^-e^+ \to \gamma X$) at
LEP is an example of direct production of unparticles.  Notice that,
beside neutrinos ($\nu \bar \nu$), $X$ can be any new hypothetical
particle, in particular, unparticle stuff.  In this case, neutrino
production is the background reaction, because the signatures for
detection of unparticles are also the missing energy and momentum.

The recent progress of neutrino physics experiments offers an
interesting scenario for studying unparticle physics. In this article
we derive bounds on unparticle physics using both neutrino-electron
scattering data coming from reactors, including the interference term
between SM amplitude.  We have derived limits from single-photon
production in electron-positron collisions.  We have also estimated
the sensitivity of upcoming neutrino-nuclei coherent scattering
measurements to unparticle physics.  Moreover, we have also compared
our results with previous works that either used the same processes
that we considered or astrophysical phenomena, and discuss the
different hypothesis that should be fulfilled for each limit; in some
cases our constraints are better than the previously reported values,
and in general they are obtained from a more detailed analysis of the
experiments under consideration and, therefore, more robust. We also
have corrected some factors derived in previous works.

Our paper is organized as follows: In Section~\ref{sec:pheno} we
review the unparticle phenomenology and derive all relevant cross
sections. The numerical results are obtained in
Section~\ref{sec:results}. Finally the discussion of the results and
our conclusions are given in Section~\ref{sec:conclusions}.

\section{Unparticle phenomenology}
\label{sec:pheno}

At energies above $\Lambda$, a hidden sector operator ${\cal O}_{UV}$
of dimension $d_{UV}$ could couple to the SM operators ${\cal
  O}_{SM}$ of dimension $d_{SM}$ via the exchange of heavy particles
of mass $M$
\begin{equation}
{\cal L}_{UV}=\frac{{\cal O}_{UV}{\cal O}_{SM}}{M^{d_{UV}+d_{SM}-4}}\,.
\end{equation}
The hidden sector becomes scale invariant at $\Lambda$ and then the
interactions become of the form
\begin{equation}
{\cal L}_{\cal U}=
C_{\cal O_U}\frac{\Lambda^{d_{UV}-d}}{M^{d_{UV}+d_{SM}-4}}\,{\cal O}_{\cal
  U}\,{\cal O}_{SM}\,,
\end{equation}
where ${\cal O}_{\cal U}$ is the unparticle operator of scaling
dimension $d$ in the low energy limit and $C_{\cal O_U}$ is a
dimensionless coupling constant.  Therefore the unparticle sector can
appear at low energies in the form of new massless fields coupled very
weakly to the SM particles.

In the low energy regime, the effective interactions for the scalar
and vector unparticle operators with the SM fermion fields are
\begin{equation}
\lambda_{0f}\frac{1}{\Lambda^{d-1}}\,\bar f f \,{\cal O_U} +
\lambda_{0\nu}^{\alpha\beta}\frac{1}{\Lambda^{d-1}}\,\bar \nu_\alpha
\nu_\beta \,{\cal O_U}
\end{equation}
and
\begin{equation}
\lambda_{1f}\frac{1}{\Lambda^{d-1}}\,\bar f \gamma_\mu f \,{\cal O}_{\cal U}^\mu +
\lambda_{1\nu}^{\alpha\beta}\frac{1}{\Lambda^{d-1}}\,\bar \nu_\alpha
\gamma_\mu \nu_\beta \,{\cal O}_{\cal U}^\mu \,,
\end{equation}
where
\begin{eqnarray}
\lambda_{if} &=& C_{{\cal O_U}^if}\frac{\Lambda^{d_{UV}}}{M^{d_{UV}+d_{SM}-4}} \,, \\
\lambda_{i\nu}^{\alpha\beta} &=& C^{\alpha\beta}_{{\cal
    O_U}^i\nu}\frac{\Lambda^{d_{UV}}}{M^{d_{UV}+d_{SM}-4}} \,,
\end{eqnarray}
with $i=0$ indicating the unparticle scalar field and $i=1$ the vector
field. We use $\alpha$ and $\beta$ to denote neutrino flavors
(including flavor changing processes) and $f=e,u,d$, for electrons,
up, and down quarks, respectively.

In the following subsections we introduce the cross sections that are
relevant for our calculations. It is useful for this purpose to use
the definitions:
\begin{equation}
g^{\alpha\beta}_{if}(d)=\frac{\lambda_{i\nu}^{\alpha\beta}\lambda_{if}}{2\sin(d\pi)}
A_d
\end{equation}
and
\begin{equation}
A_d = \frac{16\pi^{5/2}}{(2\pi)^{2d}}
\frac{\Gamma(d+1/2)}{\Gamma(d-1)\Gamma(2d)} \,.
\end{equation}

\subsection{Neutrino-electron scattering mediated by unparticles}

Neutrino-electron scattering in the context of unparticles has already
been discussed in the
literature~\cite{Montanino:2008hu,Zhou:2007zq,Balantekin:2007eg}. In
this subsection we summarize the main cross sections and we also show
some differences in our computations with the results already reported
in the literature.

The neutrino-electron cross section mediated by the scalar unparticle
is given by the expression
\begin{equation}
\label{cs:e-scalar}
\frac{d\sigma_{{\cal U}_S}}{dT}=\frac{[g^{\alpha\beta}_{0e}(d)]^2}{\Lambda^{(4d-4)}}
\frac{2^{(2d-6)}}{\pi E_\nu^2}(m_eT)^{(2d-3)}(T+2m_e)\,,
\end{equation}
where $T$ is the electron recoil energy.
Note that this cross section is twice larger than the one derived in
Ref.~\cite{Balantekin:2007eg}. We have neglected terms containing a
neutrino mass, since it is much smaller than both the electron mass
and the typical energies for the process.

An additional interference term between the SM and the unparticle
amplitude should be considered for the case of a flavor conserving
scattering ($\nu_ee^- \to \nu_ee^-$)~\cite{Zhou:2007zq}. However, for
the scalar unparticle case, this term is proportional to the neutrino
mass and, therefore, it is negligible~\cite{Balantekin:2007eg}.

For the case of a neutrino electron scattering mediated by vector
unparticles, the differential cross section has the form
\begin{equation}
\label{cs:e-vector}
\frac{d\sigma_{{\cal U}_V}}{dT}=\frac{1}{\pi}
\frac{[g_{1e}^{\alpha\beta}(d)]^2}{\Lambda^{(4d-4)}}
2^{(2d-5)}(m_e)^{(2d-3)}(T)^{(2d-4)}
\left[1 + \left(1-\frac{T}{E_\nu}\right)^2-\frac{m_eT}{E_\nu^2}\right]\,,
\end{equation}
which is 8 times larger than the cross section obtained for the same
process in Ref.~\cite{Balantekin:2007eg}. 

We would like to comment on the differences between the scalar and the
vector unparticle cross sections derived here and in
Ref.~\cite{Balantekin:2007eg}. There is a factor 4 in the vector case
due to a typo in Eq.~(12) of Ref.~\cite{Balantekin:2007eg}: the factor
$2^{(2d-8)}$ appearing there should be
$2^{(2d-6)}$~\cite{Balantekin-Ozansoy}. Another factor 2 difference in
both cross sections comes from the averaging over initial spins of
massive neutrinos~\cite{Balantekin-Ozansoy} performed in
Ref.~\cite{Balantekin:2007eg}. Here we do not average over the spins
of initial neutrinos, because the deviations from the left (right)
polarizations of initial neutrinos (antineutrinos) are highly
suppressed by the smallness of neutrino masses. 

In the neutrino-electron scattering mediated by the vector unparticles
an additional interference term should be considered for the flavor
conserving case, which is given by
\begin{equation}
\label{cs:e-vector-interf}
\frac{d\sigma_{{\cal U}_V-SM}}{dT}=\frac{\sqrt{2} G_F}{\pi} 
\frac{g_{1e}(d)}{\Lambda^{(2d-2)}}\, (2m_e T)^{(d-2)}
m_e \left\{
g_L + g_R \left(1-\frac{T}{E_\nu}\right)^2 - \frac{\left(g_L+g_R\right)}{2} 
\frac{m_e T}{E_\nu^2}
\right\} \,.
\end{equation}

This interference term for vector unparticles is linearly proportional
to the SM couplings and to the unparticle couplings, therefore it can
be bigger than the pure unparticle contribution shown in
Eq.~(\ref{cs:e-vector}) for some values of the couplings.  Note, however,
that this term would not appear in the case of flavor changing
interactions, $\nu_e e\to \nu_{\mu,\tau} e$.  In other words, neutrino
flavor conserving and neutrino flavor changing scatterings are
equivalent to the cases with and without the interference term
(\ref{cs:e-vector-interf}), respectively.

\subsection{Single-photon production in electron-positron collisions}

Direct production of an unparticle with a single-photon in
electron-positron collisions were studied in
references~\cite{Chen:2007qr,Cheung:2007zza,Cheung:2007ap}. In
reference~\cite{Chen:2007qr} there is also a prediction for unparticle
detection at the International Linear Collider (ILC).  The
differential cross section for the interaction $e^+e^-\to\gamma {\cal
U}_V$ is given by:
\begin{equation}
\frac{d\sigma_{\gamma {\cal U}}}{d\Omega}=\frac{1}{2s}\overline{|{\cal
    M}|^2}\frac{A_d}{16\pi^3\Lambda^2}\left(\frac{P_{\cal
      U}^2}{\Lambda^2}\right)^{(d-2)}E_\gamma dE_\gamma \,,
\end{equation}
with
\begin{equation}
\overline{|{\cal M}|^2}=2\lambda_{1e}^2e^2\frac{u^2+t^2+2sP_{\cal U}^2}{ut} \,,
\end{equation}
$u$, $t$, and $s$ being the Mandelstam variables.

Then the total cross section can be written as :
\begin{equation}\label{monophoton}
\frac{d\sigma_{\gamma{\cal U}}}{dx}
=\int_{y_{min}}^{y_{max}}\frac{A_d}{(4\pi)^2}
\left(\frac{\lambda_{1e}e}{\Lambda}\right)^2
\left[\frac{s(1-x)}{\Lambda^2}\right]^{(d-2)}\frac{x^2+x^2y^2+4(1-x)}{x(1-y^2)}dy
\,, 
\end{equation}
with $x=E_\gamma/E_{beam}$ and $y=\cos\theta_\gamma$, $\theta_\gamma$
being the angle between the incident beam and the outgoing photon.

\subsection{Coherent neutrino-nuclei scattering mediated by unparticles}

When momentum transfer, $Q$, is small comparing with inverse nucleus
size, $QR \leq 1$, a coherent neutrino-nucleus scattering can take
place~\cite{Freedman:1973yd}.  Since for most nuclei the typical
inverse sizes are in the range from 25 to 150~MeV, the condition for
full coherence in the neutrino-nuclei scattering is well satisfied for
reactor neutrinos and other artificial neutrino sources.

There are currently several experimental proposals that intend to
observe for the very first time this
process~\cite{Wong:2008vk,Scholberg:2005qs,Barbeau:2007qi}, while
other experimental setups have also been
studied~\cite{Bueno:2006yq,Volpe:2006in}.  The potential of some of
these experimental proposals for constraining new physics, such as
non-standard neutrino
interactions~\cite{Barranco:2005yy,Scholberg:2005qs} or a non-zero
neutrino magnetic
moment~\cite{Scholberg:2005qs,McLaughlin:2003yg,deGouvea:2006cb,Wong:2005pa}
has already been discussed.

Here we derive the coherent neutrino-nucleus scattering cross section
with intermediate scalar unparticles, in analogy with the
neutrino-electron scalar unparticle scattering cross section, and we
find:
\begin{equation}
\frac{d\sigma_{{\cal U}_S}^{\nu N}}{dT}=\frac{1}{\Lambda^{(4d-4)}} \,
\frac{2^{(2d-6)}}{\pi E_\nu^2} 
\left[g_{0u}(d)(2Z+N)+g_{0d}(d)(Z+2N)\right]^2
(m_A T)^{(2d-3)}(T+2m_A)\, ,
\end{equation}
where $T$ is the recoil energy of the entire nucleus target, $Z$ and
$N$ are the number of protons and neutrons, respectively, of the
detector nucleus target, and $A$ the mass number ($A=Z+N$). As in the
neutrino-electron scattering case, the interference term is 
proportional to the neutrino mass and can be safely neglected.

The neutrino-nucleus coherent scattering cross section mediated by
vector unparticle has the form:
\begin{eqnarray}\label{pureunparticle}
  \frac{d\sigma_{{\cal U}_V}^{\nu N}}{dT} =&&
  \frac{2^{(2d-5)}}{\pi\Lambda^{(4d-4)}}
  m_A(m_AT)^{(2d-4)}
  \left[g_{1u}(d)(2Z+N)+g_{1d}(d)(Z+2N)\right]^2 \nonumber \\ 
  &&\times \left[1  + \left(1-\frac{T}{E_\nu}\right)^2-\frac{m_AT}{E_\nu^2}\right]\,.
\end{eqnarray}
In case of the flavor conserving process the interference between SM and
vector unparticle fields is linearly proportional to the neutrino-unparticle
couplings, as we show in the following expression:
\begin{eqnarray}\label{SM-unparticle}
\frac{d\sigma_{{\cal U}_V-SM}^{\nu N}}{dT}=&&
\frac{\sqrt{2} G_F}{\pi} \frac{\left[g_{1u}(d)(2Z+N)+g_{1d}(d)(Z+2N)\right]}
{\Lambda^{(2d-2)}}2^{d-1}m_A(m_AT)^{(d-2)}\nonumber \\
&&\times \left(g_V^p Z+g_V^n N\right)\left[ 1 + \left(1-\frac{T}{E_\nu}\right)^2-\frac{m_AT}{E_\nu^2}\right] \,,
\end{eqnarray} 
where $g_V^{p,n}$ are the SM neutral current vector couplings of neutrinos with
protons $p$ and with neutrons $n$, defined as
\begin{eqnarray}
&&g_V^p=\rho_{\nu N}^{NC}\left(
\frac12-2\hat\kappa_{\nu N}\hat s_Z^2
\right)+
2\lambda^{uL}+2\lambda^{uR}+\lambda^{dL}+\lambda^{dR},\nonumber\\
&&g_V^n=-\frac12\rho_{\nu N}^{NC}+
\lambda^{uL}+\lambda^{uR}+2\lambda^{dL}+2\lambda^{dR}\,.
\label{vcouplings}
\end{eqnarray}
Here $\hat s_Z^2=\sin^2\theta_W=0.23120$, $\rho_{\nu N}^{NC}=1.0086$,
$\hat\kappa_{\nu N}=0.9978$, $\lambda^{uL}=-0.0031$,
$\lambda^{dL}=-0.0025$ and
$\lambda^{dR}=2\lambda^{uR}=7.5\times10^{-5}$ are the radiative
corrections given by the PDG~\cite{Yao:2006px}.  In order to obtain
both the SM as well as the interference term,
Eq.~(\ref{SM-unparticle}), we have neglected the axial contribution
since the ratio of the axial to the vector contributions is expected
to be of the order $1/A$, $A$ being the atomic number. We have also
considered the axial and vector form factors equal to unity, which
is a good approximation for $Q^2 \ll m_A^2$, where $Q$ is the
transferred momentum.

\section{Analysis and results}
\label{sec:results}

With the cross sections obtained in the previous section, it is
possible to obtain constraints on different unparticle parameters from
the experimental data presented in the literature. In this section we
report the constraints that we have derived from a $\chi^2$ analysis
applied to the relevant experiments.

\subsection{Neutrino-electron scattering}\label{elastic}

We performed an analysis of the $\bar\nu_e e \to \bar \nu e$
considering the MUNU data. In order to estimate the constraints on the
parameters
$d$ and $\lambda_0$ ($\lambda_1$) $ = \sqrt{\lambda^{e\beta}_{0\nu}\lambda_{0e}}$
($\sqrt{\lambda^{e\beta}_{1\nu}\lambda_{1e}}$)
we compute the integral
\begin{equation}
\sigma =  \int dT' \int dT \int dE_\nu 
                    \frac{d\sigma_{{\cal U}_{S,V}}}{dT} \lambda (E_\nu) R(T,T')
\label{diff:cross:sec-sm}
\end{equation}
with $R(T,T')$ the energy resolution function for the MUNU
detector. The relative energy resolution in this detector was
found to be 8\% and it scales with the power $0.7$ of the
energy~\cite{Daraktchieva:2005kn}.

We use an anti-neutrino energy spectrum $\lambda (E_\nu)$ given by
\begin{equation}
\lambda(E_{\nu}) = \sum_{k=1}^4 a_k \lambda_k(E_\nu) \,,
\label{diff:cross:sec-NSI}
\end{equation}
where $a_k$ are the abundances of $^{235}$U ($k=1$), $^{239}$Pu
($k=2$), $^{241}$Pu ($k=3$), and $^{238}$U ($k=4$) in the
reactor; $\lambda_k(E_\nu)$ is the corresponding neutrino energy
spectrum which we take from the parametrization given
in~\cite{Huber:2004xh}, with the appropriate fuel composition. For
energies below $2$~MeV there are only theoretical calculations for the
antineutrino spectrum which we take from Ref.~\cite{Kopeikin:1997ve}.

\begin{figure}
        \centering
        \includegraphics[width=0.8\columnwidth]{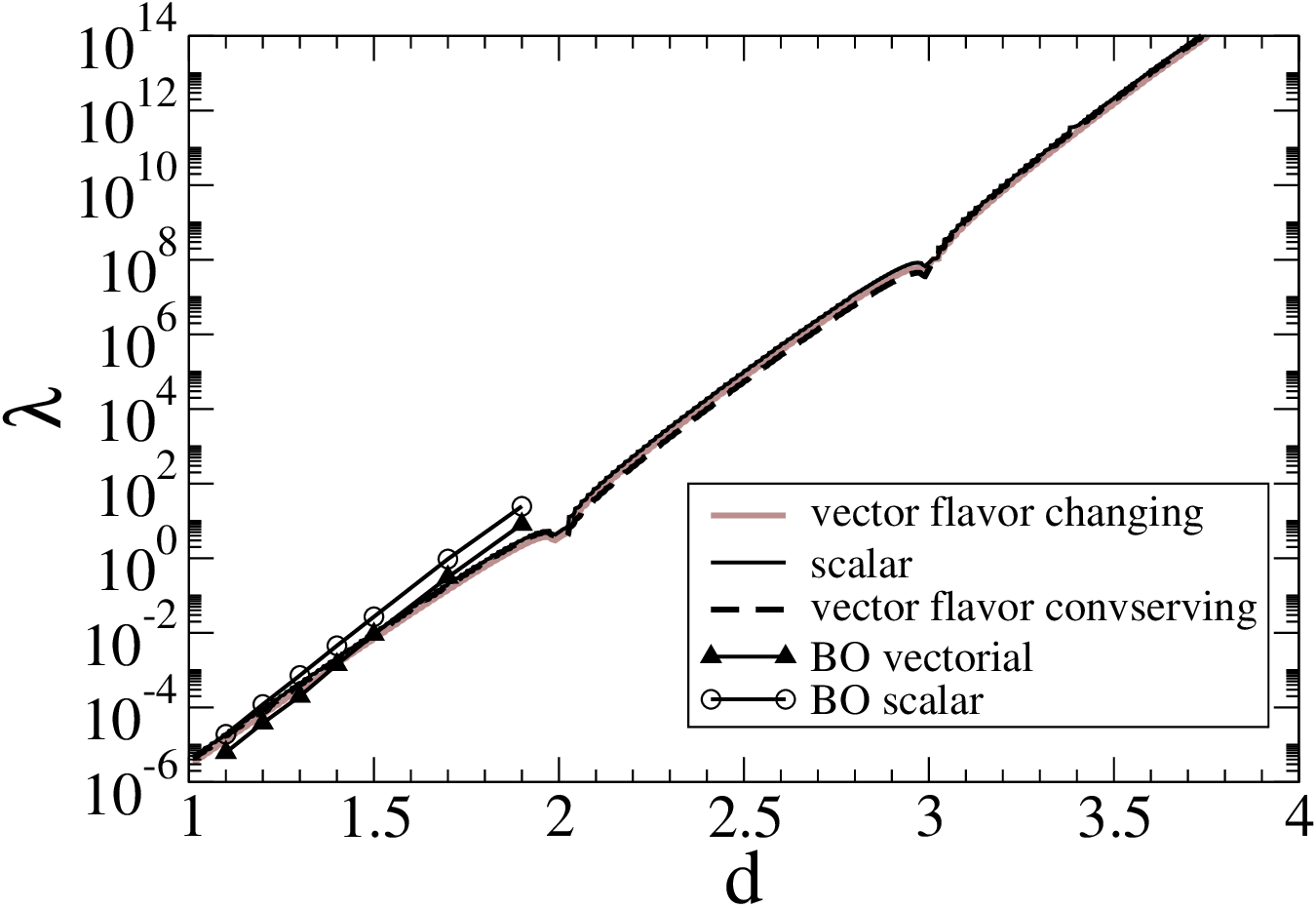}
\caption{Limits on the parameters $d$ and
  $\lambda_{0,1}=\sqrt{\lambda^{e\beta}_{0,1\nu}\lambda_{0,1e}}$ (90
  \% CL) from the MUNU experiment for the scalar unparticle case
  (black solid line) and for the vector unparticle cases, both for
  flavor changing currents (grey solid line) and for the flavor
  conserving conserving case (dashed line). Previous bounds obtained
  by Balantekin and Ozansoy (BO)~\cite{Balantekin:2007eg} (dots and
  triangles) are shown for comparison. The present analysis based
  on the MUNU data gives stronger constraints on $\lambda_{0,1}$ for
  values of $d>1.5$.}
\label{fig:nu-e}
\end{figure}

With this formula we can compute the number of events expected in MUNU
in the case of a SM cross section, as well as in the case of an extra
contribution due to unparticle physics, for the parameters $d$ and
$\lambda_0$. We are considering $\Lambda=1$~TeV. 

The expected number of events in the case of an unparticle
contribution to the neutrino-electron scattering $N_i^{\rm
theo}=N_i^{\rm SM}+N_i^{{\cal U}_{S,V}}$, can be compared with
measured number of events per day, $N^{\rm exp} = (1.07 \pm 0.34)$
events/day, reported by the MUNU
collaboration~\cite{Daraktchieva:2005kn}, We show the results of our
analysis in Fig.~\ref{fig:nu-e}, where the maximum allowed
values of the unparticle parameters are shown at 90\% C.L.  We also
show in the same plot the results obtained in previous
analysis~\cite{Balantekin:2007eg}.

The same analysis was done for the vectorial case and the result 
is  shown in the same Fig.~\ref{fig:nu-e}. We show both the
result that considers the interference term ($\nu_e e \to \nu_e e$) as
well as the case where such interference term is absent ($\nu_{e} e \to \nu_{\mu ,\tau} e$) . Finally, we also show previous
reported results from Ref.~\cite{Balantekin:2007eg} for comparison.

\begin{figure}
        \centering
        \includegraphics[angle=0,width=0.8\columnwidth]{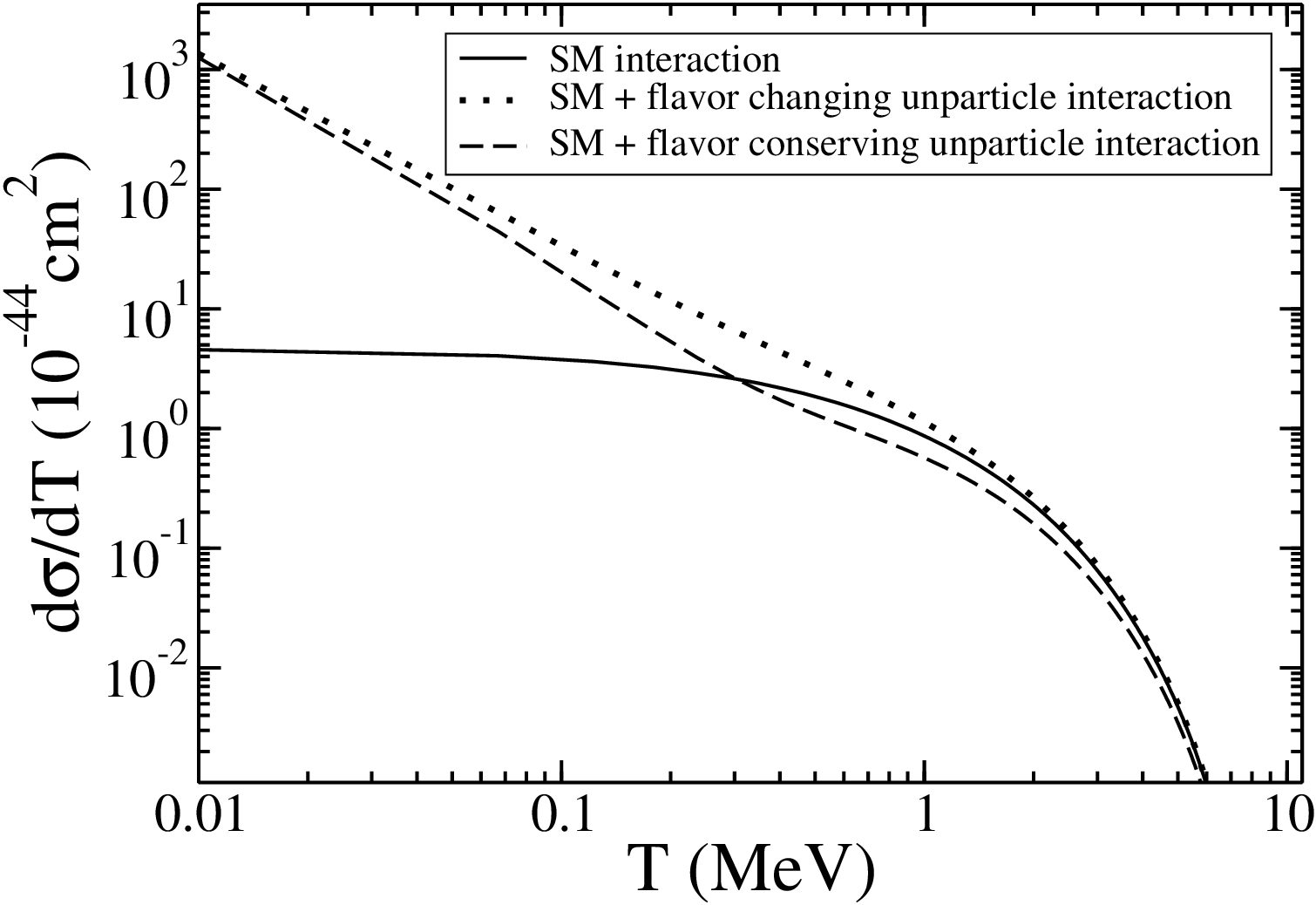}
\caption{Differential cross section for $\nu$-$e$ scattering for the
  SM case and for the vector unparticle case. We show both  the
  flavor conserving as well as the flavor changing case.  In the
  flavor conserving interaction mediated by unparticles, the negative
  interference term gives a different spectral shape.
  The effective coupling
  $\lambda_1=\sqrt{\lambda_{1\nu}^{\alpha\beta}\lambda_{1e}}$ was fixed
  to $\lambda_1=5.5 \times 10^{-5}$ and $d=1.2$. }
\label{nu_ecrosssection}
\end{figure}

In order to illustrate the sensitivity of future neutrino electron
scattering experiments and to show the behavior of the different
unparticle interactions, we show in Fig.~\ref{nu_ecrosssection} the
differential cross section antineutrino scattering off electrons.
Several experimental proposals plan to perform an accurate measurement
of this process~\cite{Deniz:2008rw,Volpe:2006in,Adams:2008cm}. It is clear
from this figure that besides the increase in the expected number of
events, the shape of the spectrum will also change in different energy
regions.

\subsection{Limits from single-photon production with unparticles}

The real emission of unparticle plus a single photon in
electron-positron collisions at LEP has the same signature of missing
energy carried by neutrino pairs plus single-photon production.

The best data on single-photon production plus missing energy has been
collected by the four LEP experiments: ALEPH, DELPHI, L3 and
OPAL~\cite{Abdallah:2003np,lep,Achard:2003tx}.
We analyze this data considering the sum of the cross sections for
single-photon production with neutrino pairs and unparticle.

Disagreements between our calculations and the Monte Carlo results
quoted by the LEP collaborations are included as an additional
theoretical uncertainty which we have added in quadrature in the
calculation of our errors~\cite{Barranco:2007ej}. Because of the small
systematic error they have, we can assume that all of them are
independent, with no correlation between them.  In the case of the
more recent DELPHI data analysis~\cite{Abdallah:2003np}, we perform
our analysis considering the cross section reported, instead of the
number of events.

\begin{figure}
        \centering
        \includegraphics[width=0.8\columnwidth]{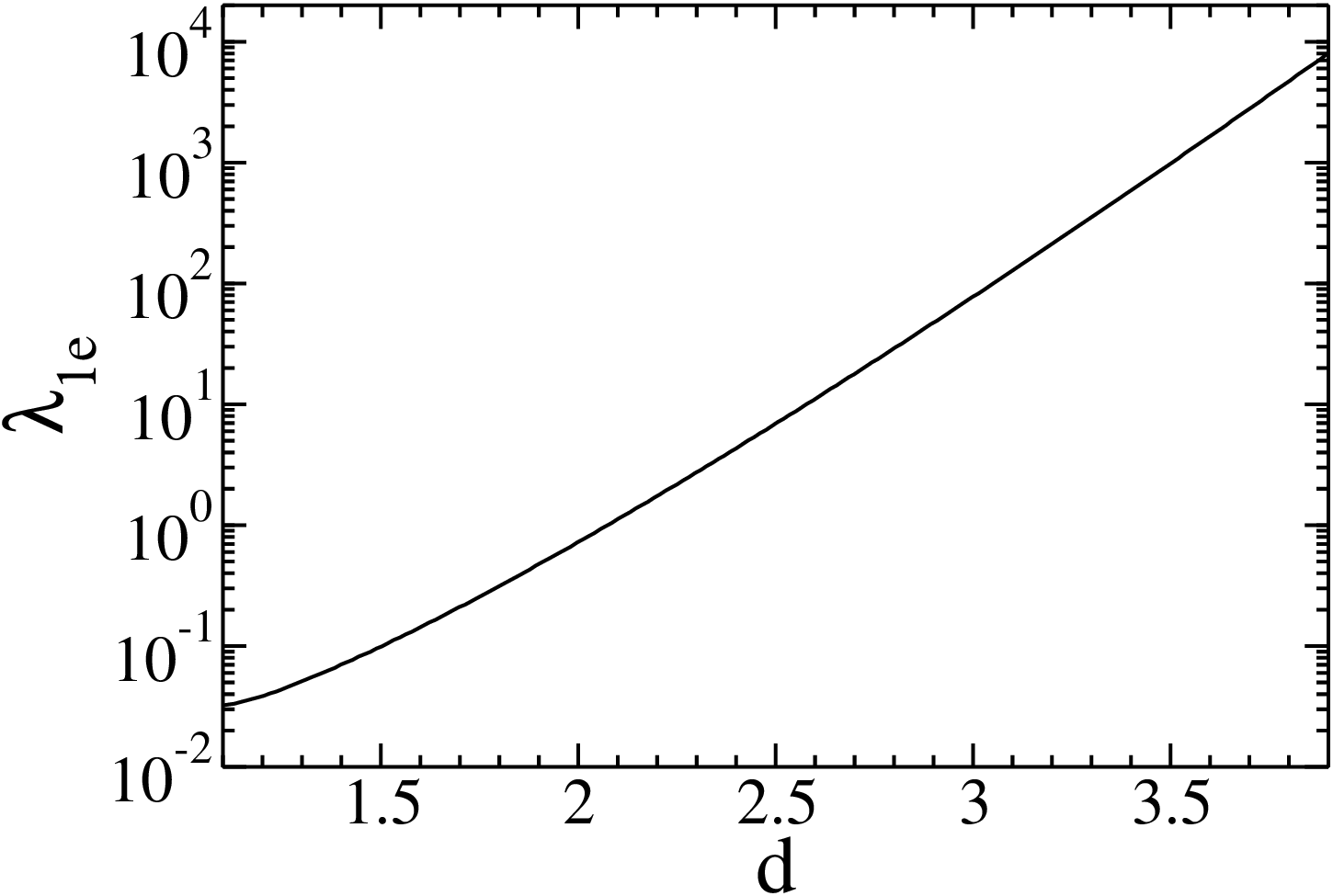}
    \caption{Limits on the parameters $d$ and $\lambda_{1e}$ for the
    unparticle analysis of the four LEP experiments at 90\% C.L. considering
    $\Lambda=1~$TeV.}
\label{fig:lep}
\end{figure}

The results of our analysis are presented in Fig.~\ref{fig:lep} and in
Table~\ref{collidercheung}. In Table~\ref{collidercheung} we show the
comparison of our results with the previous results obtained in
Ref.~\cite{Cheung:2007jb} and we also compare these results with a
possible future limit that can be obtained with ILC for a center of
mass energy, $\sqrt{s}=500$~GeV.

The analysis made in Ref.~\cite{Cheung:2007jb} considered the cross
section limit of $\sigma \sim 0.2~$pb at 95\% C.L. for the process
$e^+e^-\to\gamma X$ obtained by L3~\cite{Achard:2003tx} under the cuts
$E_\gamma>5$~GeV, $|\cos\theta_\gamma|< 0.97$, and $\sqrt{s}=207$~GeV.
By fixing the coupling $\lambda_{1e}=1$, bounds on the energy scale
$\Lambda$ are obtained for different values of $d$.  Our limits are
looser but more robust in the sense that we have used all LEP
experiments data and obtained the constraints from a $\chi^2$
statistical analysis. ILC limits would be stronger for large $d$'s,
i.e., for $d>1.8$.
\begin{table}
\begin{tabular}{c|c|c|c}
\hline
$d$ & $\Lambda$ (TeV) from \cite{Cheung:2007jb} & $\Lambda$ (TeV) Our analysis & $\Lambda$ (TeV) Future ILC \\
\hline
2.0 & 1.35 & 1.1 & 1.69 \\ 
1.8 &4 & 3.1 & 4.25 \\
1.6 & 23 & 22.1 & 17.9 \\
1.4 & 660  & 612 & 257 \\
\hline
\end{tabular}
\caption{Limits on $\Lambda$ from single-photon production data of
$\sigma(e^+e^- \to \gamma \mathcal{U}_V)$ from LEP data,
$\lambda_{1e}=1$, $95\%$ C.L. In the last column we show possible
future bounds for a center of mass energy of
$\sqrt{s}=500$~GeV.}\label{collidercheung}
\end{table}
 
\subsection{Sensitivity of coherent neutrino-nucleus scattering}

\begin{figure}
\includegraphics[angle=0,width=0.8\textwidth]{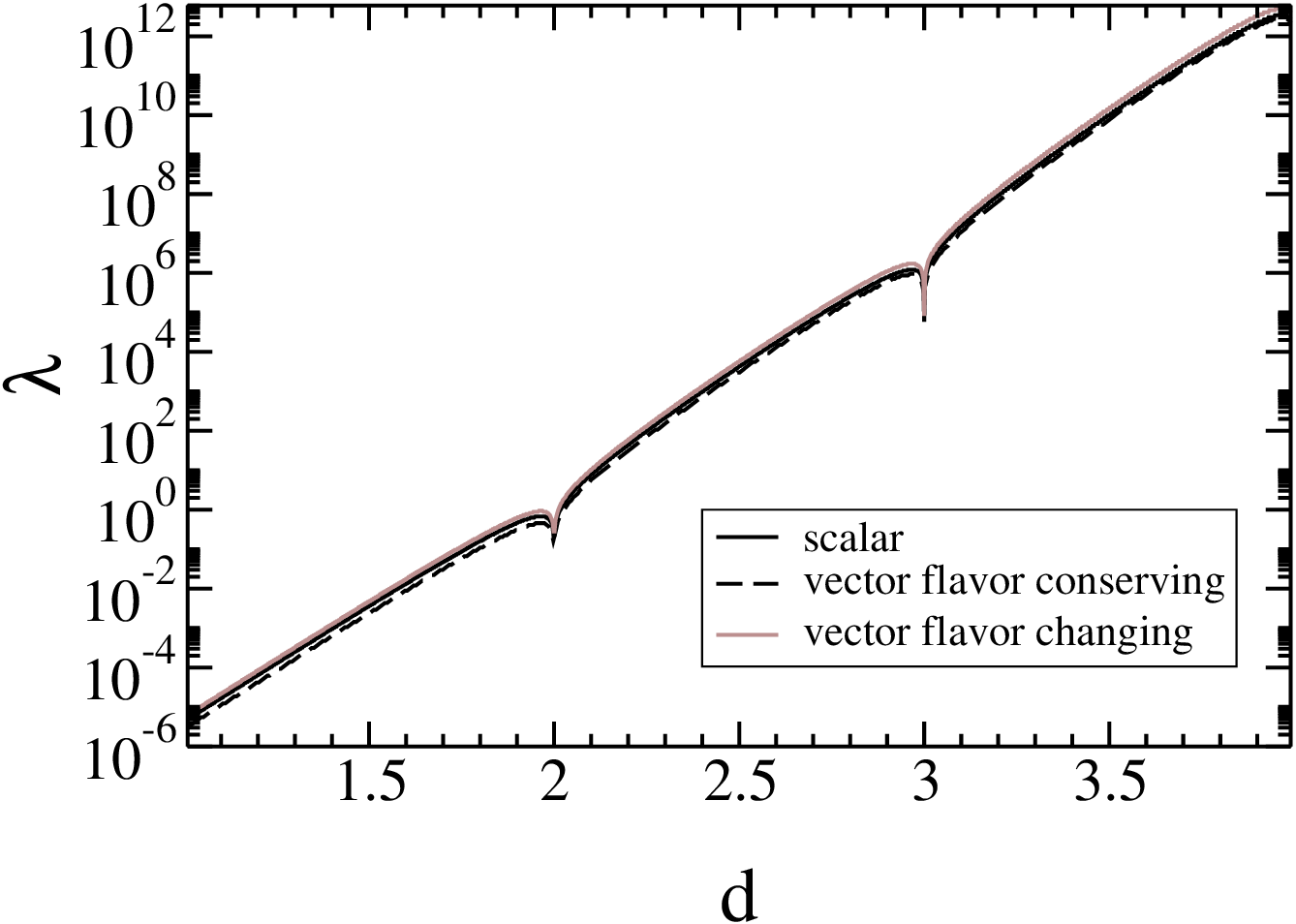}
\caption{Future sensitivity of the TEXONO proposal (90\% C.L.) on the unparticle 
         dimension $d$ and the effective coupling $\lambda$.
         Scalar case corresponds to $\lambda=\sqrt{\lambda^{e\beta}_{0\nu}\lambda_{0d}}$ (black solid line).
         Vector flavor conserving for $\lambda=\sqrt{\lambda^{ee}_{1\nu}\lambda_{1d}}$ (dashed line) and
         vector flavor changing $\lambda=\sqrt{\lambda^{e\beta}_{1\nu}\lambda_{1d}},~\beta = \mu,\tau$
         (grey solid line). Limits  were done assuming $\lambda_{0,1u}=0$. The flavor conserving case,
         which includes the interference term, is the most
         sensitive. }\label{Texono}
\end{figure}

The coherent neutrino-nuclei scattering can be a great complementary
tool in order to constrain physics beyond the standard model such as
unparticle physics. As already mentioned, there are several
experimental proposals that intend to observe this
process~\cite{Wong:2008vk,Scholberg:2005qs,Barbeau:2007qi}. To show
the sensitivity of such experiments to unparticle parameters we
consider for definiteness the TEXONO collaboration proposal which has
started a program towards the measurement of the coherent $\nu-N$
scattering by using reactor neutrinos and 1~kg of an
``ultra-low-energy'' germanium detector (ULEGe)~\cite{Wong:2008vk}.
The number of expected events, neglecting for the moment the detector
efficiency and resolution, can be calculated by:
\begin{equation}
N_{\rm{events}}^{\rm SM}=t\phi_0\frac{M_{\rm{detector}}}{m_A}
\int\limits_{E_{min}}^{E_{max}}dE_\nu
\int\limits_{T_{th}}^{T_{max}(E_\nu)}dT
\lambda(E_\nu)\frac{d\sigma^{\nu N}_{\rm SM}}{dT}(E_\nu,T)\,,
\label{Nevents}
\end{equation}
with $t$ being the period of data acquisition, $\phi_0$ the total neutrino
flux, $M_{\rm{detector}}$ the total mass of the detector,
$\lambda(E_\nu)$ the normalized neutrino spectrum, $E_{max}$ the
maximum neutrino energy and $T_{th}$ the detector energy
threshold. The maximum nucleus' recoil energy depends
on the nucleus mass $m_A$ through the relation
\[
T^{max}=2(E_\nu^{max})^2/(m_A+2E_\nu^{max})\,.
\]  
For the TEXONO proposal we take a minimum threshold energy
of $T_{th}=400~$eV. We have estimated the sensitivity for the TEXONO
proposal to constrain unparticle parameters by means of a $\chi^2$
analysis
\begin{equation}
\chi^2={\left(N_{\rm{events}}^{\rm{SM}}-N_{\rm{events}}^{{\cal U}_{S,V}}\right)^2
\over \delta N_{\rm events}^2} \,,
\end{equation}
where we have calculated $N_{\rm{events}}^{{\cal U}_{S,V}}$ exchanging
the SM differential cross section in Eq.~(\ref{Nevents}) by the cross
section given in Eqs.~(\ref{cs:e-scalar}) and (\ref{cs:e-vector}), for
the scalar and vectorial unparticles respectively.  In
Fig.~\ref{Texono} we show the sensitivity of the coherent
$\nu-N$ scattering for the scalar unparticle propagator. We shown 
also the sensitivity for the case when the propagator has a vectorial
structure. As we have discussed, in this case there is an interference 
between the scattering mediated by the vectorial unparticle propagator and the
usual SM scattering mediated by the $Z$ boson. We can see that the sensitivity
becomes more stringent when this interference is included.  
In all the previous cases we have fixed the scale $\Lambda=1$~TeV.

\section{Discussion and summary}
\label{sec:conclusions}

\begin{figure}
        \centering
        \includegraphics[angle=0,width=0.8\columnwidth]{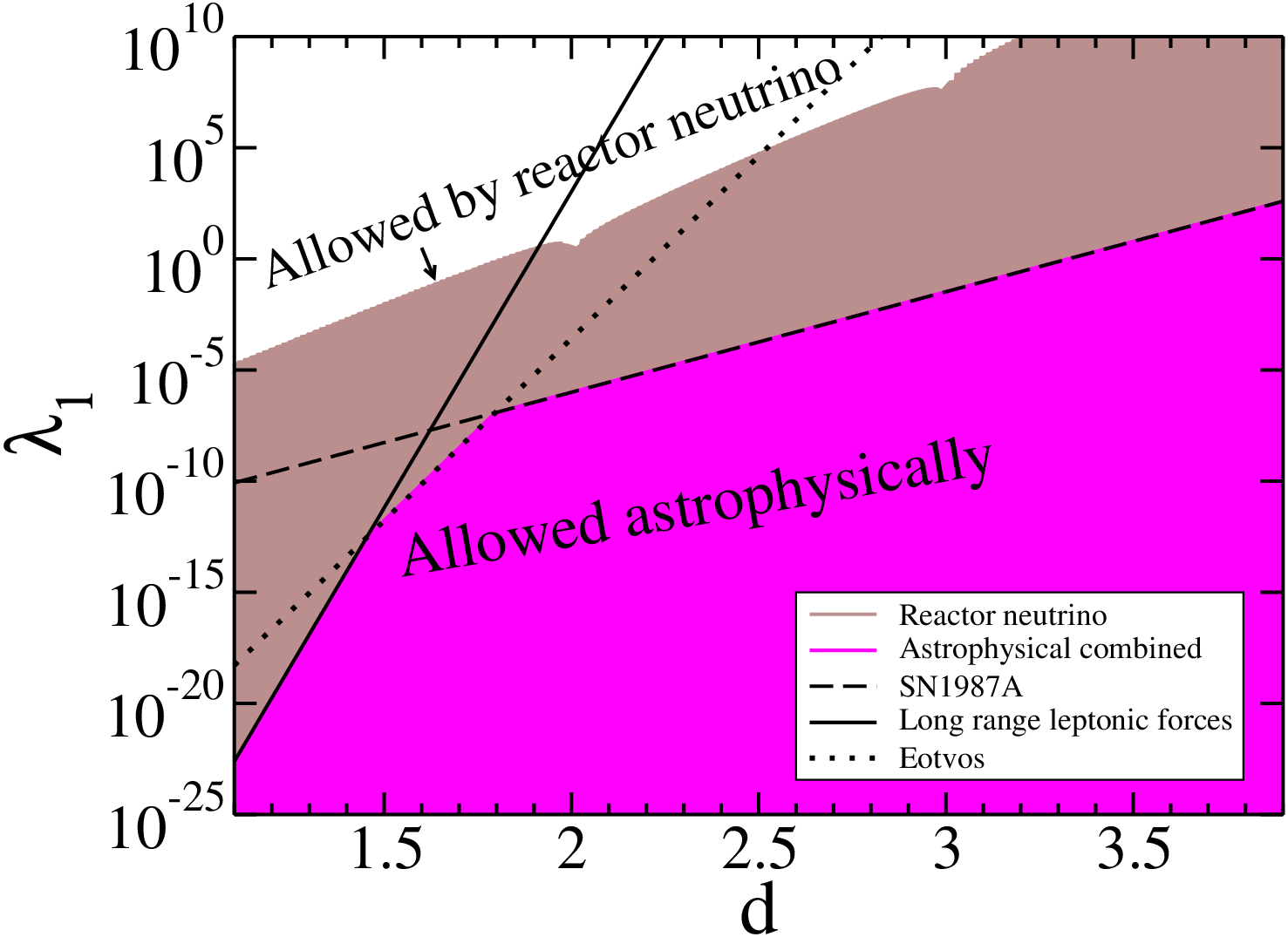}
\caption{Current constraints on vectorial unparticle couplings
 $\lambda_1$ and $d$ from reactor neutrino $\nu-e$ elastic scattering
 (MUNU experiment). The current astrophysical limits are shown for
 comparison, although both for the SN1987A and for E\"otv\"os case different 
 initial hypothesis should be considered (see text for details.)
}
\label{astrophysical}
\end{figure}

So far, in section \ref{sec:results}, we have derived the current
bounds on the relevant unparticle's parameter by using the current
available neutrino data from reactor and from LEP experiments. We have
also shown the future sensitivity for coherent neutrino-nucleus
scattering. There are, however, other limits obtained from
astrophysical observations.  We would like to discuss three of them,
namely, from the observation of supernova SN1987A
neutrinos~\cite{Davoudiasl:2007jr,Hannestad:2007ys}, from the tests of
gravitational inverse square law (E\"otv\"os-type or fifth force
experiments)~\cite{Deshpande:2007mf,Freitas:2007ip} and the limits
obtained by the possible existence of new electronic long range
forces. We will emphasize that, despite these limits are much stronger
than those coming from reactor and accelerator experiments, they are
valid under certain assumptions and therefore the terrestrial limits
shown here give an important complementarity.

The limits obtained from neutrinos coming from SN1987A in
Refs.~\cite{Davoudiasl:2007jr,Hannestad:2007ys} were derived under the
assumption that unparticles could freely escape supernova core, thus
releasing a large amount of energy and therefore leading to a decrease
of the duration of the neutrino burst during supernovae
explosion. However, if the couplings are large enough this could
cause trapping of unparticles in the supernova due to their
interaction with the dense medium in the core, which therefore would
relax the present
constraints~\cite{Davoudiasl:2007jr,Hannestad:2007ys}.  

Other very strong constraints on unparticle interactions with the SM
particles were obtained from experiments testing the Newtonian law of
gravity~\cite{Deshpande:2007mf,Freitas:2007ip} and positronium
decays~\cite{Liao:2007bx}.  However, if the theory is not exactly
scale invariant, or if scale invariance is broken at some scale
smaller than a millimeter, thereby screening the long range forces,
then these limits will not
apply~\cite{Deshpande:2007mf,Freitas:2007ip,Rajaraman:2008qt}. Therefore,
although we will consider in what follows values that are bigger than
these constraints, they may well be allowed under the appropriate
assumptions.

Finally, let us comment on the possibility that long-range forces
could be originated by unparticles. In~\cite{GonzalezGarcia:2008wk} it
was found that solar neutrino data can constrain the vector and scalar
unparticle interactions. The constraints obtained in
\cite{GonzalezGarcia:2008wk} can be re-written as
\begin{equation}\label{conchaboundscalar}
(\lambda_{0\nu}^{ee}-\lambda_{0\nu}^{\nu a})\left(\lambda_{0e}+\lambda_{0p}+
\lambda_{0n}\left\langle \frac{Y_n}{Y_e}\right\rangle\right)
\frac{\Gamma(d+1/2)\Gamma(d-1/2)}
{2 \pi^{2d}\Gamma(2d)(R_{\odot}\Lambda)^{2(d-1)}}
 < 6.8\times 10^{-45} 
\end{equation}
\begin{equation}\label{conchaboundvector}
(\lambda_{1\nu}^{ee}-\lambda_{1\nu}^{\nu a})\left(\lambda_{1e}+\lambda_{1p}+
\lambda_{1n}\left\langle \frac{Y_n}{Y_e}\right\rangle\right)
\frac{ \Gamma(d+1/2)\Gamma(d-1/2)}
{2 \pi^{2d}\Gamma(2d)(R_{\odot}\Lambda)^{2(d-1)})}  < 4.5\times 10^{-53}
\end{equation}
where $\lambda_{0,1\nu}^{\nu
  a}=\mbox{cos}^2\theta_{23}\lambda_{0,1\nu}^{\mu\mu}+
\mbox{sin}^2\theta_{23}\lambda_{0,1\nu}^{\tau\tau}$, $\theta_{23}$ is
the solar mixing angle. $Y_{n,e}$ are the relative number densities of
neutrons and electrons respectively and $\langle ...\rangle$ means
average along the neutrino trajectory. Bounds
(\ref{conchaboundscalar}) and (\ref{conchaboundvector}) are given at
$3\sigma$ C.L.

Let us assume for the moment, and just as an illustrative example,
that $\lambda_{0,1\nu}^{\nu a}=\lambda_{0p}=\lambda_{0n}=0$. In this
particular case we can see that the constraints
(\ref{conchaboundscalar}-\ref{conchaboundvector}) involve the same
parameters that our parameter $\lambda$ shown in
Fig.~(\ref{fig:nu-e}).  We have also plotted this constraint
(\ref{conchaboundvector}) in Fig. \ref{astrophysical} and we can see
that, for this special case, indeed long-range forces are very
restrictive for values of $d$ close to one, while for $d > 2$ reactor
neutrinos are more restrictive.

In Fig.~\ref{astrophysical} and Table~\ref{tablavectorial} we report
our limits on $\lambda_1$ for the vectorial unparticle case obtained
by using the MUNU neutrino data (Section~\ref{elastic}).  For the
E\"otv\"os-type limit we have closely followed
Ref.~\cite{Deshpande:2007mf} with a different interpolation on
$\beta_k$.  Instead of a linear interpolation, we interpolated
$\beta_k$ as a function of $1/k$ and $1/k^2$ for the values reported
in \cite{Adelberger:2006dh}. $k$ is related with the unparticle
parameter dimension $d$ through the relation
$k-1=2d-2$~\cite{Deshpande:2007mf}.  The Long Range force limits where
obtained from Eq.~(\ref{conchaboundvector}) and for the limits from
supernova cooling (SN1987A) we have used the limit obtained in
Ref.~\cite{Hannestad:2007ys}. Finally, we also show in the same
table, the limits reported in~\cite{Montanino:2008hu}, that were
obtained by considering the recent Borexino data; please note that in 
this case the reported limits apply to the scalar coupling, but we 
show them for the sake of completeness.
\begin{table}[!t]
    \begin{center}
        \begin{tabular}{c|c|c|c|c|c}
            \hline
            d  & $\nu-e$ scattering & E\"otv\"os & Long range & SN1987A &
Solar $\nu$'s  \\
            \hline
            \hline
            1.1 &$2.0\times 10^{-5}$&$6.3\times 10^{-19}$&$2.8\times
10^{-23}$
&$9.1\times 10^{-11}$ & $1.1\times 10^{-5}$\\
            1.25 &$1.9\times 10^{-4}$&$1.6\times 10^{-16}$&$5.2\times
10^{-19}$&
$4.0\times 10^{-10}$ & $1.2\times 10^{-4}$\\
            1.5 &$9.7\times 10^{-3}$&$1.7\times 10^{-12}$&$5.7\times
10^{-12}$&
$5.7\times 10^{-9}$ & $7.3\times 10^{-3}$\\
            1.75 &$3.7\times 10^{-1}$&$2.6\times 10^{-8}$&$6.1\times
10^{-5}$&
$7.4\times 10^{-8}$ & $3.4\times 10^{-1}$ \\
            2.1 &$40.$&$1.1\times 10^{-2}$&$6.0\times 10^{5}$&
$2.9\times 10^{-6}$ & $100.$ \\
            2.25 &$713$&$4.2$&$1.0\times 10^{10}$&
$1.3\times 10^{-5}$ & $1127.$\\
            2.5 &$5.5\times 10^{4}$&$4.8\times 10^{4}$&$1.1\times 10^{17}$&
$1.8\times 10^{-4}$ & $6.6\times 10^{4}$\\
            2.75 &$2.9\times 10^{6}$&$5.5\times 10^{8}$&$1.8\times 10^{24}$&
$2.3\times 10^{-3}$ & $3.5\times 10^{6}$ \\
            3.1 &$1.2\times 10^{9}$&$3.3\times 10^{14}$&$1.1\times 10^{34}$&
$9.9\times 10^{-2}$ & $1.0\times 10^{9}$\\
            3.25 &$2.3\times 10^{10}$&$9.6\times 10^{16}$&$3.1\times
10^{38}$&
$4.7\times 10^{-1}$ & $1.1\times 10^{10}$\\
            3.5 &$2.1\times 10^{12}$&$1.5\times 10^{21}$&$3.2\times
10^{45}$&
$6.1$ & $6.7\times 10^{11}$\\
            3.75 &$1.1\times 10^{14}$&$1.9\times 10^{25}$&$3.3\times
10^{52}$&
$87.2$ & $3.5\times 10^{13}$\\
            3.9 &$1.1\times 10^{15}$&$6.2\times 10^{27}$&$5.8\times
10^{56}$&
$414.3$ & $4.0\times 10^{14}$\\
            \hline
        \end{tabular}
    \end{center}
    \vskip -0.2cm
    \caption{Constraints on the vector coupling $\lambda_1$ from the
 neutrino electron scattering experiment, and from astrophysical
 limits.  The confidence level considered in different reported
 results is different and therefore the comparison is
 qualitative. Besides, for the SN1987A and for E\"otv\"os case
 different initial hypothesis should be considered (see text for
 details). For the sake of completeness, we show in the last column
 limits coming from solar data for the case of the $\lambda_0$ scalar
 coupling.}
    \label{tablavectorial}
\end{table}

We can summarize now the results shown in this work as follows:  
\begin{itemize}
\item 
we have corrected the neutrino-electron cross-sections and calculated
the coherent neutrino-nucleus scattering cross-sections for the
unparticle case.
\item 
we have obtained the constraints on unparticle couplings with
neutrinos and electrons coming from available reactor and accelerator
experiments, specifically MUNU and LEP data.
\item 
we have included into the analysis the interference term for the
vector unparticle case of flavor conserving scattering and we have 
shown its relevance.
\item 
we have compared our results with astrophysical limits and have
discussed that, although the astrophysical constraints are stronger
than the direct experimental bounds, they are based on some
assumptions which may be violated and, therefore, both type of limits
are relevant and complementary.
\begin{itemize}
\item 
we have found that reactor limits are stronger than E\"otv\"os-type
(fifth force) limits for values of $d>2.55$, and stronger than the
long-range leptonic force limits for values of $d>1.95$.
SN1987A limits are always stronger than the reactor limits.
\end{itemize}
\item
We have obtained LEP limits derived from accounting for all four LEP
experiments and the sensitivity of ILC is also given.
\item
we have estimated future sensitivity of coherent netrino scattering
experiments to the neutrino-quark unparticle interaction.
\end{itemize}

\begin{acknowledgments}
  We would like to thank Baha Balantekin, Okan Ozansoy and Georg Raffelt for
  the fruitful discussions.  This work has been supported by CONACyT,
  SNI-Mexico and PAPIIT grant IN104208.  TIR was supported by the DFG
  (Germany) under the grant SFB/TR 27.  TIR thanks Physics Department of
  CINVESTAV for the hospitality during the visit when part of this work was
  done.
\end{acknowledgments}


\end{document}